\documentstyle[aps,prl]{revtex}

\begin{document}

\input epsf.sty

\twocolumn[\hsize\textwidth\columnwidth\hsize\csname %
@twocolumnfalse\endcsname

\title{Magnetic Behavior of the Cuprate Superconductors}
\author{Alexander Sokol}
\address{%
Department of Physics, University of Illinois at Urbana-Champaign,
Urbana IL 61801 \\
and L.D. Landau Institute for Theoretical Physics, Moscow, Russia}

\date{cond-mat/9505070, May 16, 1995}

\maketitle

\begin{abstract}

I review recent work
\protect\cite{CSY,Sokol:Pines,Sachdev:SS,Sokol:LANL,SCS,Barzykin:Pines}
on magnetic dynamics of
the high temperature superconductors using a model that combines
two weakly interacting
species of low-energy excitations: the antiferromagnetic
spin waves which carry spin-1 and no
charge, and Fermi-liquid-like quasiparticles which carry spin-$1/2$ and
charge $e$. The model allows conversion of spin waves into electron-hole
pairs; however, the low-energy spin waves are not collective modes of the
quasiparticles near the Fermi surface,
but rather are a separate branch of the low-energy spectrum.
With certain experimentally justified
assumptions, this theory is remarkably universal: the dependence on
the detailed microscopic Hamiltonian and on doping can be absorbed
into several experimentally measurable parameters.
The $z=1$ theory of the insulators and $z=2$ theory of the
overdoped materials, are both reproduced as limiting cases of the
theory described here, which predicts that the
underdoped materials remain in $z=1$ universality class at sufficiently
high temperature.
This theory provides a framework for understanding both the
experimental results and microscopic calculations, and in particular
yields a possible explanation of the spin gap
phenomenon. I also discuss some of the
important unresolved issues.

\bigskip

To appear in the Proceedings of the Stanford Conference on
Spectroscopies in Novel Superconductors (1995).
\end{abstract}


\vspace{0.2in}

\phantom{.}
]

\section{Introduction}

\begin{figure}
\centerline{\epsfxsize=3.0in\epsfbox{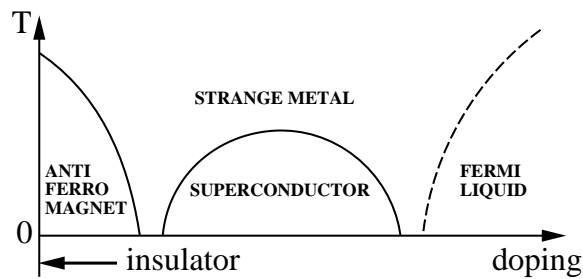}}
\bigskip
\caption{
A phase diagram of the high-temperature superconductors as a function of
doping and temperature.
In the limit of zero doping, the insulating parent materials
develop a long-range N\'eel order and their low-energy excitation
spectrum has only one species of excitations --- the spin waves. The
overdoped materials at low temperatures exhibit Fermi-liquid-like
properties; there, the quasiparticles are the only species of low-energy
excitations. We attempt to understand the
properties of the intermediate ``strange metal'' phase, where
superconductivity occurs,
by combining the excitations found in the extreme
doping limits, namely the spin waves and the quasiparticles, and allowing
spin waves to convert into electron-hole pairs.
}
\label{fig:strangemetal}
\end{figure}

The evolution of the normal state properties of the high temperature
superconductors with doping
is schematically  described by the phase diagram
in Fig.\ref{fig:strangemetal}.
The stoichiometric insulators $\rm La_2CuO_4$ and $\rm YBa_2Cu_3O_6$
undergo an
antiferromagnetic transition at $T_N$, and exhibit short-range
antiferromagnetic correlations well above $T_N$.
Upon doping by more than several per cent strontium or oxygen, these systems
become metallic and no longer have a long-range antiferromagnetic
order, but short-range magnetic correlations remain.
It is universally agreed that the
normal state properties of this metallic state are far from
the conventional metallic behavior,
and the term ``strange metal'' has been
coined to describe them. Some of the key properties of the ``strange
metal'' are: (i) short-range antiferromagnetic correlations,
seen by NMR and neutron scattering; (ii) both the resistivity and
the uniform magnetic susceptibility increase as the temperature
increases; (iii) some, but not all, of the materials exhibit a suppression
of the low-frequency spectral weight for magnetic excitations for
$T<T^*\sim 150K$ (spin gap). At higher doping, the normal state
properties become increasingly similar to the Fermi
liquid (FL) behavior with short-range
antiferromagnetic correlations. While
Landau Fermi liquid theory in the orthodox sense requires
temperature-independent spin correlations, the temperature dependence is
weak in $\rm YBa_2Cu_3O_7$
and other fully doped materials, where close proximity
of the FL state is evident.

High-temperature superconductivity occurs in the intermediate strange
metal phase in Fig.\ref{fig:strangemetal}.
Currently, there is no consensus regarding the precise
microscopic Hamiltonian of this phase.
In what follows, I describe
a theoretical approach to this problem
which bypasses the missing information about the microscopic
Hamiltonian by assuming continuous evolution of the
low-energy magnetic and charge excitation spectrum between the
antiferromagnetic insulator and the Fermi liquid phases, through the
intermediate strange metal phase
(see, e.g.
\cite{CSY,Sokol:Pines,Sachdev:SS,Sokol:LANL,SCS,Barzykin:Pines}, and
references therein, and relevant earlier work
\cite{CHN,Shraiman:Siggia,Millis:Monien:Pines,Millis}).

\section{Nearly Antiferromagnetic Fermi Liquid}

The approach of
Refs.\cite{CSY,Sokol:Pines,Sachdev:SS,Sokol:LANL,SCS,Barzykin:Pines}
is based on
combining the low-energy excitations encountered
in the two extreme doping limits in Fig.\ref{fig:strangemetal}: the
undoped antiferromagnetic insulators where
the low-energy excitations are spin waves, and the overdoped limit
where the low-energy excitations are the usual Fermi liquid
quasiparticles. I emphasize that such an approach
makes no explicit assumptions about the microscopic Hamiltonian,
and also it
makes no assumptions regarding the mechanism through which the two types
of low energy excitations are formed. Instead, the theory is based on
the premise that
at low energies, the spectral weight of the system is {\em
shared} between the spin waves and the quasiparticles,
and that the primary interaction mechanism is spin wave to electron-hole
pair conversion. A microscopic basis for this theory is provided by
the mean-field spin density wave (SDW) calculations,
described in \cite{SCS}.

The presence of the first species of
low-energy excitations, the quasiparticles,
is evident from photoemission experiments.
Such quasiparticles at low energies must be
similar to the quasiparticles of the Landau Fermi liquid theory, and
carry both charge $e$ and spin $S=1/2$, in order
to account for the sharp
quasiparticle peak near the Fermi energy, observed in
photoemission. The quasiparticles are
gapless and form a Fermi surface, the shape of which is very important for
the interaction between the quasiparticles and
spin waves.

The other species of low energy excitations, the spin waves, are
the only low-energy modes in the antiferromagnetic insulators
$\rm La_2CuO_4$ and
$\rm YBa_2Cu_3O_6$. Below $T_N$, the spin waves are
gapless, and in small-$q$ limit have a linear spectrum
$\omega_q\simeq cq$ and a mean-free path that is longer than the wavelength.
When elevated temperature or doping
destroy the long-range order, and only short-range antiferromagnetic
correlations remain (the correlation length $\xi$ is finite),
the spin waves are no longer
gapless and also are overdamped for $q<\xi^{-1}$. Nevertheless,
for larger wavevectors $q>\xi^{-1}$, the spin waves
remain nearly
the same as in the presence of long-range order, because they
sample only the local order for distances of order
wavelength $\lambda=2\pi/q\ll \xi$, and therefore are nearly
insensitive to the absence of antiferromagnetic
correlations at distances larger than
$\xi$. The evidence for their existence in the strange metal
phase is based primarily on NMR and neutron scattering measurements,
and is discussed in what follows.
The spin waves carry spin $S=1$ and no charge.

Whereas on the microscopic level an antiferromagnetic
spin wave can be regarded as a coherent electron-hole pair,
the contribution from the quasiparticle states
near the Fermi surface accounts for only a
fraction of the total spectral weight of the spin wave,
if there is no nesting. The spin waves
are therefore a separate branch of excitations, rather than a
collective mode of low-energy quasiparticles.
In that sense, the theory described here is somewhat similar to the
spin-charge separation theory, where low-energy spin excitations
cannot be represented as collective modes of low-energy charge
excitations, even though on the microscopic level both originate from
electrons. The difference between our model
\cite{CSY,Sokol:Pines,Sachdev:SS,Sokol:LANL,SCS,Barzykin:Pines},
and the spin-charge separation picture applied to the high temperature
superconductors \cite{Anderson}, is that in our model
one species of excitations
(spin waves) carries only spin, while the other
(quasiparticles) carries both spin and charge, therefore
there is charge separation, but not spin separation.
Recently, Sachdev \cite{Sachdev:SS}
and Laughlin \cite{Laughlin} have pointed
out that in a system with spin-charge separation,
some of the spinon and holon excitations
may form bound states which carry both spin and charge, and
in many respects are similar to conventional quasiparticles.

\begin{figure}
\centerline{\epsfxsize=3.0in\epsfbox{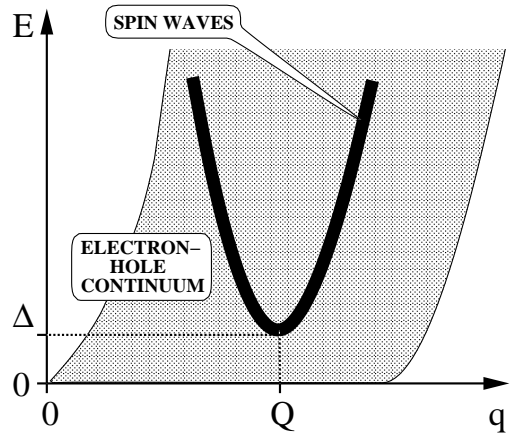}}
\bigskip
\caption{
Magnetic excitation spectrum of type (B) model
(classification due to Millis \protect\cite{Millis}), where ${\bf
Q}=(\pi/a,\pi/a)$ can connect two points on the Fermi surface.
The minimum of the spin wave dispersion $\omega_q=(c^2|{\bf q}-{\bf
Q}|^2+\Delta^2)^{1/2}$ is located inside the
electron-hole continuum.
}
\label{fig:continuum}
\end{figure}

One of the key factors that affect the magnetic and charge properties is
the position of the  minimum
of the spin wave spectrum, ${\bf Q}=(\pi/a,\pi/a)$ for the commensurate
short range order discussed here, with respect to the
electron-hole continuum. If the spin wave spectrum
is located inside the electron-hole continuum, as it is illustrated in
Fig.\ref{fig:continuum}, the spin waves
can convert into electron-hole pairs. This conversion process
becomes the dominant contribution to spin wave damping at low energies,
and has a profound effect on the magnetic properties. In the
classification introduced by Millis, this situation is called model
B. For a discussion of the opposite case, when ${\bf Q}$ cannot
connect two points on the Fermi surface  (model A),
see Refs.\cite{Millis,Sachdev:SS,SCS}.

The results of recent photoemission measurements
\cite{photoemission} indicate that model
B is relevant to the Y- and Bi-based high-temperature
superconductors. In what follows, I concentrate
exclusively on model B.

\section{Spin Gap}

The spin gap behavior is the
suppression of the low frequency spectral weight for magnetic excitations,
which can be observed by NMR and neutron scattering.
In the model at hand, this
suppression is expected already in a very crude approximation which
treats spin wave and quasiparticle excitations
as noninteracting, free modes.
Below, I describe a qualitative picture of the spin gap behavior
suggested by Sokol and Pines \cite{Sokol:Pines,Sokol:LANL}, and
derived starting from model B by Sachdev, Chubukov, and Sokol
\cite{SCS}.

In the non-interacting approximation, the
density of states for magnetic excitations is a sum of spin wave and
quasiparticle contributions. For frequencies below
the gap for spin wave excitations, $\omega<\Delta=c/\xi$, only the
quasiparticles contribute to the magnetic density of states. For
frequencies $\omega>\Delta$, both the spin waves and the quasiparticles
contribute. In a measurement of the total magnetic density of states, for
example in a NMR relaxation or  neutron scattering measurement at low
temperatures, one expects to
see a gap-like suppression of the density of states for magnetic
excitations for $\omega<\Delta$.

Beyond the non-interacting approximation,
the lifetime of spin waves primarily is due to their conversion
into electron-hole pairs.  Upon increasing doping, the spin gap
is washed out by a combination of two separate effects: first,
the spin waves become overdamped because the rate of conversion
increases; second, the gap for spin-wave
excitations $\Delta=c/\xi$ increases as $\xi$ decreases.
The remaining spin waves, which are pushed to much larger energies, can in
principle be seen by neutron scattering.
At least two recent experiments seem to allow such an
interpretation: the evidence for the magnetic character of the
$41\,$meV
peak in $\rm YBa_2Cu_3O_7$ by Keimer {\em et al.} \cite{Keimer}
and the observation of the zone-boundary magnon in
$\rm La_{2-x}Sr_xCuO_4$
by Aeppli {\em et al.} \cite{Aeppli}.
Keimer \cite{Keimer} and Barzykin and Pines \cite{Barzykin:Pines} have
suggested that the $41$meV feature
is a coherent, triplet, electron-hole pair -- in other words, a spin wave
(several alternative interpretations have been suggested as well). The zone
boundary high-energy spin wave (a magnon), which continuously
evolved from the identical mode in the insulator, seems to be most
likely explanation of Aeppli's measurements; however, this data is very
preliminary \cite{Aeppli}.

\section{Universal Theory of Magnetic Dynamics}

In this section, I describe recent work by Sachdev, Chubukov, and
Sokol \cite{SCS}, where the universal
behavior of the magnetic correlations near ${\bf Q}$, and of the bulk
susceptibility, was calculated using the rate of the Landau
damping of spin waves as an input parameter. Several
important assumptions about the microscopic model are
built into this theory. First, we limit our study to systems with
short-range order at $T=0$. The region at low doping where the
long range N\'eel order exists at $T=0$ requires separate
consideration.  Second, we require that
there is no nesting, i.e.
the areas of the Fermi surface that are adjacent to the points connected
by ${\bf Q}$, are not parallel to each other. The photoemission
measurements \cite{photoemission}
show that this is the case in Y- and Bi-based
superconductors. Third, we assume that there is no substantial
$q$-,  $\omega-$, and $T-$dependence of the Landau damping for
$|{\bf q}-{\bf Q}|\sim 1/\xi$ and $\omega\sim\Delta$, respectively.
This assumption is more difficult to verify experimentally; it holds
in mean field theory on the disordered side if the
correlation length is large enough.

Under these assumptions, the additional damping due to the spin wave
to electron-hole pair conversion can be included into the theory by
inserting its rate $\Gamma$ into the
noninteracting spin wave dynamical response function (a similar
expression was introduced by Barzykin, Pines, Sokol, and Thelen
\cite{BPST} on phenomenological grounds):
\begin{equation}
\chi({\bf q},\omega) \sim \frac{\mbox{const}}{\omega_q^2 - \omega^2} \to
\frac{\mbox{const}}{\omega_q^2 - \omega^2 - i\omega\Gamma},
\label{chi:bare}
\end{equation}
where the spin wave spectrum for finite $\xi$ (short-range
correlations) has the following form:
\begin{equation}
\omega_q = \sqrt{c^2 |{\bf q}-{\bf Q}|^2 + \Delta^2}, \ \ \ \
\Delta=c/\xi.
\end{equation}
The exact dynamical susceptibility of the model
is obtained by calculating the
effects of mutual scattering of spin waves, primarily the additional
damping, on $\chi({\bf q},\omega)$
given by Eq.(\ref{chi:bare}). When the Landau
damping processes dominate dissipation,
the result of such a calculation
differs only slightly from Eq.(\ref{chi:bare}).

With these assumptions, the dynamical magnetic susceptibility near the
staggered wavevector, $|{\bf q}-{\bf Q}|\ll a^{-1}$, is given by:
\begin{equation}
\chi ({\bf q}, \omega ) = \frac{{\cal Z}}{T^{-\eta}}
\left( \frac{c}{T} \right)^2 \Phi_s
\left(\frac{c|{\bf q}-{\bf Q}|}{T},\frac{\omega}{T},\frac{\Delta}{T},
\frac{\Gamma}{T} \right),
\label{chis}
\end{equation}
and the temperature-dependent part of the uniform magnetic
susceptibility, $\chi_u$, by:
\begin{equation}
\chi_u (T) - \chi_u (T=0) = (1 + \alpha^{\prime} )\frac{T}{c^2}
\Phi_{u} \left( \frac{\Delta}{T}, \frac{\Gamma}{T}  \right).
\label{chiu}
\end{equation}
Here, $\alpha^{\prime}$ is a non-universal constant,
both $\Phi_s$ and $\Phi_u$ are universal and computable
functions of their arguments, and $\eta$ is a universal critical exponent
which is very small and for all practical purposes can be replaced by zero.

Eqs.(\ref{chis},\ref{chiu}) apply
when all dimensionful parameters and
variables are smaller than the respective lattice cutoffs, which
roughly translates into the following:
\begin{equation}
\omega,T,\Delta,\Gamma\ll \min(J,E_F),\ \ \ \ \
\xi^{-1},|{\bf q}-{\bf Q}|\ll a^{-1},
\label{restrict}
\end{equation}
where $J$ is the exchange constant, $E_F$ the Fermi energy, and
$a$ the lattice spacing. If the conditions (\ref{restrict}) apply, the
magnetic dynamics near ${\bf Q}$
depends on four dimensionful parameters, two of
which describe the spin wave spectrum ($c$ and $\Delta$), one is set
by the size of spin ($\cal Z$), and one is determined by the rate
of the Landau damping ($\Gamma$). The temperature-dependent part of the
bulk susceptibility, which adds to the Pauli term,
is also universal upto the overall prefactor $1+\alpha'$.

For a low-energy theory
this number of parameters is not excessive because they
include {\em all} the information about the material that is
reflected in its magnetic behavior. Three of them
are the overall scales of magnetic moment, distance, and energy, and are
therefore unavoidable in any low-energy theory;
determination of these three
parameters from the experimental data is straightforward and
is a matter of simple normalization.
The fourth parameter serves as the single crossover variable which
describes the evolution from the scaling behavior of the insulators
(the dynamical exponent $z=1$) to that of the overdoped materials
($z=2$). One of the key results of \cite{SCS}
is that there is only {\em one} such crossover parameter.
Here the dynamical exponent $z$ relates
temperature dependences of the characteristic frequency scale
for magnetic correlations,
$\bar{\omega}$, and the characteristic length scale (the correlation
length, $\xi$):
\begin{equation}
\omega \sim \xi^{-z}.
\label{z}
\end{equation}

\begin{figure}
\centerline{\epsfxsize=3.0in\epsfbox{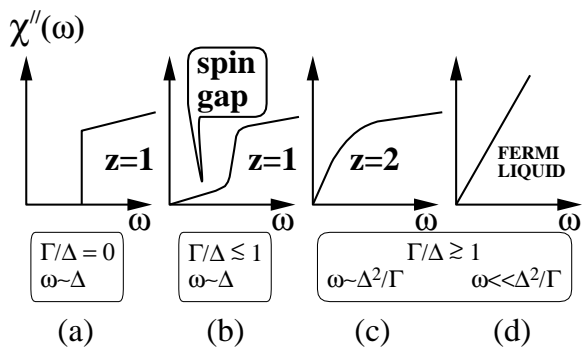}}
\bigskip
\caption{Evolution of $\chi_L''({\bf q},\omega) = \int d{\bf q}\,
\chi''({\bf q},\omega)$ at $T=0$
as a function of the dimensionless crossover
parameter $\Gamma/\Delta$.
The underdoped materials $\rm YBa_2Cu_3O_{6.63}$ and
$\rm YBa_2Cu_4O_8$ are in regime (b), and have $z=1$.
The same applies to $\rm La_{1.85}Sr_{0.15}CuO_4$ at high
temperatures; at low temperatures,
some modifications of the theory due to the incommensurability of
short-range order are necessary. The fully doped material
$\rm YBa_2Cu_3O_7$ is in regime (c), and has $z=2$.}
\label{fig:evolution}
\end{figure}

Fig.\ref{fig:evolution}
illustrates the evolution of zero temperature magnetic properties
with doping on the example of the local dynamical magnetic susceptibility:
\begin{equation}
\chi_L''(\omega) = \int d{\bf q}\, \chi''({\bf q},\omega),
\end{equation}
which shows a gradual
crossover, observed as a function of $\Gamma/\Delta$,
from $z=1$ regime ($\Gamma/\Delta\alt 1$ \cite{CSY,Sokol:Pines,SCS}) to
$z=2$ regime ($\Gamma/\Delta\agt 1$ \cite{Millis:Monien:Pines,Millis}).
The crossover between
$z=1$ and $z=2$ regimes at low temperature
is expected, and observed experimentally,
as doping increases; in the universal theory, a similar crossover is also
expected at a fixed doping as the temperature increases; it is however
not observed experimentally because scaling fails altogether when at
high temperatures the correlation length becomes very short,
$\xi/a\alt 2$ (Barzykin and Pines \cite{Barzykin:Pines}).
The $z=2$ phase  shows a further crossover to the Landau
Fermi liquid when $\Delta^2/\Gamma$ becomes larger than $\omega$ or $T$.
Fig.\ref{fig:evolution} describes the sequence of
crossovers observed as $\Gamma/\Delta$ increases.

\subsection{z=1 limit ($\Gamma/\Delta \protect\alt 1$)}

The limit
$\Gamma/\Delta=0$ and $\Delta>0$ corresponds to an insulator with
short-range antiferromagnetic correlations at $T=0$
(Fig.\ref{fig:evolution}a).
Such an insulator
has a true energy gap $\Delta$ because creating any combination of
excitations (spin waves which have a gap) above the ground state
requires a finite energy, hence $\chi_L''(\omega<\Delta)=0$. While this
limit has no direct relevance to the cuprate oxide insulators (which
develop N\'eel long-range order at $T=0$ and have gapless spin
waves), it helps to understand the behavior of the underdoped
materials where $\Gamma/\Delta\alt 1$. The effect of small doping is
shown in Fig.\ref{fig:evolution}b: $\chi_L''(\omega)$ becomes finite for
all $\omega$, but it remains much smaller for $\omega<\Delta$ compared
to $\omega>\Delta$. The gap which existed for $\Gamma/\Delta=0$
transforms into a knee-like feature at $\omega\sim\Delta$
for $\Gamma/\Delta\alt 1$. This
behavior reproduces the essential features of
the spin gap observed in some of the underdoped
high-temperature superconductors.

At high temperatures $T\gg \Delta$, the behavior of $\Gamma/\Delta\alt
1$ underdoped system is similar to that of the  $\Gamma/\Delta=0$
insulator, because for energies $\omega\agt \Delta$ the spectral
weight of the two systems is nearly the same, and in both cases
is dominated by the spin wave contribution
(compare Figs.\ref{fig:evolution}a,b).
As a result, the magnetic dynamics in this temperature range is in the
$z=1$ universality class, and exhibits quantum critical behavior
$\bar{\omega}\sim \Delta \sim T$ \cite{CSY,CHN}.

The NMR measurements in the underdoped materials
$\rm YBa_2Cu_3O_{6.63}$ (Takigawa \cite{Takigawa6}) and
$\rm YBa_2Cu_4O_8$ (Imai {\em et al.}
\cite{124T2-Imai}, Stern {\em et al.}
\cite{124T2-Stern}, and Corey {\em et al.} \cite{124T2-Corey})
are consistent with $z=1$ predictions by Sokol and
Pines \cite{Sokol:Pines}: both materials have
a spin gap at low temperatures, and $T_1T/T_{\rm 2G}\approx
\mbox{const}$ at high temperatures. In
$\rm La_{1.85}Sr_{0.15}CuO_4$, $1/T_{\rm 2G}$ has not yet been
measured at sufficiently high temperatures ($T=500-1000K$), but
$1/T_1$ has been, and it
is nearly the same as in the insulator
(Imai {\em et al.}\cite{214T1-Imai}).
Therefore, it has to be dominated by the spin wave contribution, which
obeys the $z=1$ theory.
At low temperatures, the application of this theory
to $\rm La_{1.85}Sr_{0.15}CuO_4$ requires modifications
due to the incommensurability of short range order.
According to the theory
described here, $\rm La_{1.85}Sr_{0.15}CuO_4$ should have a smaller
spin gap than the underdoped YBCO materials, which appears to be
almost obscured by the superconducting phase (see
Ref.\cite{Barzykin:Pines} for a discussion).

\subsection{z=2 limit ($\Gamma/\Delta \protect\agt 1$)}

Figs.\ref{fig:evolution}c,d describe different regions of the
same frequency dependence
of $\chi_L''(\omega)$ at $T=0$ in the overdoped case, where
$\Gamma/\Delta\agt 1$:
(c) corresponds to $\omega\sim \Delta^2/\Gamma$, and (d) to
$\omega\ll \Delta^2/\Gamma$. Regime (c) is $z=2$ quantum critical;
regime (d) is $z=2$ quantum disordered, which is
equivalent to the Landau Fermi liquid.
The theory of magnetic behavior in the $z=2$ regime has been developed by
Millis, Monien, and Pines \cite{Millis:Monien:Pines}
and Millis \cite{Millis}.

For temperatures $T\gg \Delta^2/\Gamma$,
the characteristic frequency scale for magnetic correlations
$\bar{\omega}\sim T$ modulo $\log(\Delta/T)$ corrections, and
the correlation length is given by Eq.(\ref{z}) with $z=2$.
In the opposite limit
$T\ll\Delta^2/\Gamma$, the correlation length is nearly
T-independent, as is expected for the Landau Fermi liquid.

The fully doped material $\rm YBa_2Cu_3O_7$ is in the $z=2$
universality class, and its behavior between $T_c$ and room
temperature is intermediate between the
$z=2$ quantum critical regime
(Fig.\ref{fig:evolution}c) and $z=2$ quantum disordered regime
(Fig.\ref{fig:evolution}d). The NMR experiments find
$T_1T/T_{\rm 2G}^2\approx \mbox{const}$ in agreement with the
$z=2$ quantum
critical prediction (Imai {\em et al.} \cite{Imai7});
however, both of the rates do not vary strongly
with temperature, which is an indication of the proximity of the Fermi
liquid phase, where they saturate. This material does not have a spin
gap, in agreement with its assignment as being in the $z=2$ regime.

\section{Uniform Susceptibility}

The approach of Ref.\cite{SCS} allows one to calculate the
temperature-dependent part of the uniform spin susceptibility,
$\chi_u(T)-\chi_u(T=0)$, and correctly predicts that it monotonically
increases as a function of $T$ above $T_c$,
that its slope becomes
smaller as doping increases, and that
temperature-independent Pauli
$\chi_u(T)$ is recovered in the overdoped case.

\begin{figure}
\centerline{\epsfxsize=2.4in\epsfbox{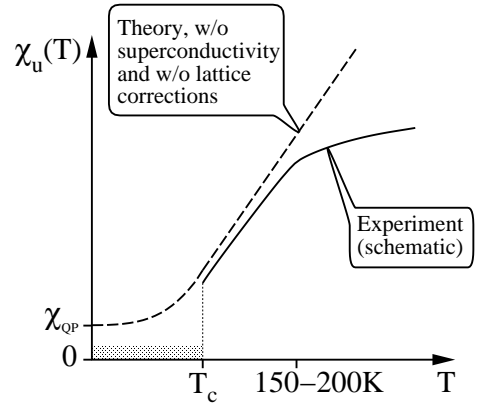}}
\bigskip
\caption{A sketch of the uniform spin susceptibility
$\chi_u(T)$ in the underdoped YBCO (solid line), and
the theoretical prediction of Eq.(\protect\ref{chiu})
(dashed line). Since Eq.(\protect\ref{chiu}) is only valid in the
normal state, comparisons below $T_c$ are not meaningful. Above $T_c$,
the experimentally observed $\chi_u(T)$ increases rapidly, and its
slope is qualitatively consistent with that predicted by
Eq.(\protect\ref{chiu}). At higher temperatures,
the theory predicts continuing linear increase,
whereas the experimental data shows saturation at
$T\sim150-200K$. Following the analogy with similar behavior of
$\chi_u(T)$ for the insulating antiferromagnets at $T\sim 0.7-0.9J$,
one may speculate that the discrepancy is due to
lattice corrections.}
\label{fig:chi}
\end{figure}

However, the experimentally observed
slope of $\chi_u(T)$ sharply decreases above
$T\sim 150-200K$ compared to the slope below this temperature but
above $T_c$ \cite{Johnston,Takigawa6,124T2-Stern},
whereas the universal theory predicts that the increase of
$\chi_u(T)$ continues with the same slope
(Fig.\ref{fig:chi}).
One may speculate that this deviation is due to lattice corrections;
detailed calculations must be performed to determine whether or not
this speculation is correct.

Millis, Ioffe, and Monien \cite{Millis:Ioffe:Monien} recently
presented a scenario of spin gap formation based on spin-charge
separation, accompanied
by singlet pairing of
spin-carrying excitations (spinons).
Since holons do not have spin and therefore do
not contribute to the magnetic density of states,
the bulk susceptibility must vanish
upon singlet pairing of spinons at low temperatures.
They argued that the observed
decrease of the bulk susceptibility at low temperatures in
underdoped YBCO
can only be explained with this scenario.

However, their conclusion relies on the assumption that both
$\chi_L''(\omega\to0,T)$ and $\chi_0(T)$ would extrapolate to zero
at low temperatures even without the superconducting transition,  a
statement that
cannot be verified experimentally. In our view \cite{SCS},
no conclusion can be drawn from
the experimental data on $\chi_u(T)$
above $T_c$ as to whether at $T\to0$ the
magnetic density of states
without superconductivity would decrease all the way down to zero, as it is
required by their model, or would  become much smaller than at high
temperatures, but remain finite, as it is expected in our model.

\section{Unresolved Issues and Challenges}

The universal scaling theory yields a qualitative scenario for the
evolution of magnetic behavior with doping which is consistent with
the experiment. It also has been very
successful in quantitatively explaining
as well as predicting the results of some of
the experimental measurements, notably the NMR relaxation rates in the
insulators and doped materials. However, it fails to explain the
detailed temperature dependence of $\chi_u(T)$ at high temperatures.
This failure is likely to be caused by the lattice
corrections in the broad sense:
the influence of the Brillouin zone boundary
on q-integrations, the non-linearity of the spin-wave spectrum at
large wavevectors, bilayer coupling in YBCO,
or energy dependence of the Landau damping rate $\Gamma$.

Different quantities are differently affected by the lattice
corrections, and some are more robust than others. For instance,
over a range of temperatures, $\chi_u$ in the Heisenberg model
is affected by the lattice
corrections, while the characteristic frequency $\bar{\omega}$ is not.
Furthermore, some of the
lattice corrections that do appear,
can be absorbed into the dimensionful
parameters of the model,
while the universal scaling functions remain nearly unaffected
\cite{EGSS,Sandvik:Scalapino}.
A detailed study of these lattice effects may allow their inclusion
into a phenomenological extension of the scaling theory to
expand its range of applicability.

Another important challenge is to determine the precise role of bilayer
coupling, which is central to the scenario of spin gap
formation by Millis and Monien
\cite{Millis:Monien}. In the theory described here, the
bilayer coupling affects the dimensionful parameters of the model,
but does not affect any observables expressed as a function of these
parameters, unless the size of bilayer
coupling is comparable to other low-energy scales. Note that
in our model, the spin gap may form with or without bilayer
coupling, but it is enhanced if bilayer coupling is present.

Finally, it is very important to
understand on the microscopic level how
the spin waves and the quasiparticles share the spectral weight for
intermediate doping, and in particular the mechanism of the
experimentally observed reduction of
the Pauli contribution to the uniform spin susceptibility in the
underdoped case.
The experimental data indicates that the spin wave to
electron-hole pair conversion rate is also suppressed in the
underdoped materials, which in the theory described here results in
spin gap formation. The cause of such a reduction is not fully understood,
and may be a consequence of a reduced density of states for
the quasiparticles at low doping, which is also seen in
$\chi_u(T=0)$.

Studies of microscopic models for the high-Tc cuprates should help to clarify
whether the experimentally observed evolution of the key
parameters of the model discussed here is similar to that obtained in
microscopic calculations, thereby allowing one to check the validity of the
proposed theory.

\section{Acknowledgments}

I would like to thank my collaborators
Andrey V. Chubukov and Subir Sachdev for many comments and
suggestions;
Victor Barzykin and David Pines for informing me of their
work \cite{Barzykin:Pines} before publication;
and S. Chakravarty, R.B. Laughlin, A.J. Millis,
D. Pines, and S. Sondhi for helpful discussions.
The author was supported by an
Alfred P. Sloan Research Fellowship during the preparation of
this manuscript.


\end{document}